\documentclass[aps,prl,twocolumn,superscriptaddress]{revtex4}
\usepackage{graphicx}
\usepackage{mathrsfs}
\usepackage{bm}
\usepackage{amsmath}
\usepackage{dcolumn}
\usepackage{epstopdf}
\usepackage{dsfont}
\usepackage{amssymb}
\usepackage{tabularx}
\usepackage{array}
\usepackage{float}
\usepackage{siunitx}
\usepackage{colordvi}
\usepackage{color}
\begin{document}
\title{Valley-Current Splitter in Minimally Twisted Bilayer Graphene}
\author{Tao Hou}
\affiliation{ICQD, Hefei National Laboratory for Physical Sciences at Microscale, University of Science and Technology of China, Hefei, Anhui 230026, China}
\affiliation{CAS Key Laboratory of Strongly-Coupled Quantum Matter Physics, and Department of Physics, University of Science and Technology of China, Hefei, Anhui 230026, China}
\author{Yafei Ren}
\affiliation{ICQD, Hefei National Laboratory for Physical Sciences at Microscale, University of Science and Technology of China, Hefei, Anhui 230026, China}
\affiliation{CAS Key Laboratory of Strongly-Coupled Quantum Matter Physics, and Department of Physics, University of Science and Technology of China, Hefei, Anhui 230026, China}
\author{Yujie Quan}
\affiliation{ICQD, Hefei National Laboratory for Physical Sciences at Microscale, University of Science and Technology of China, Hefei, Anhui 230026, China}
\affiliation{CAS Key Laboratory of Strongly-Coupled Quantum Matter Physics, and Department of Physics, University of Science and Technology of China, Hefei, Anhui 230026, China}
\author{Jeil Jung}
\affiliation{Department of Physics, University of Seoul, Seoul 02504, South Korea}
\author{Wei Ren}
\affiliation{International Centre for Quantum and Molecular Structures,\\
Materials Genome Institute, Shanghai Key Laboratory of High Temperature Superconductors, Physics Department, Shanghai University, Shanghai 200444, China}
\author{Zhenhua Qiao}
\email[Corresponding author:~~]{qiao@ustc.edu.cn}
\affiliation{ICQD, Hefei National Laboratory for Physical Sciences at Microscale, University of Science and Technology of China, Hefei, Anhui 230026, China}
\affiliation{CAS Key Laboratory of Strongly-Coupled Quantum Matter Physics, and Department of Physics, University of Science and Technology of China, Hefei, Anhui 230026, China}
\date{\today}

\begin{abstract}
  We study the electronic transport properties at the intersection of three topological zero-lines 
  as the elementary current partition node that arises in minimally twisted bilayer graphene. 
  Unlike the partition laws of two intersecting zero-lines, we find that (i) the incoming current can be partitioned into both left-right adjacent topological channels and that (ii) the forward-propagating current is nonzero. 
  By tuning the Fermi energy from the charge-neutrality point to a band edge, the currents partitioned into the three outgoing channels become nearly equal. 
  Moreover, we find that current partition node can be designed as a perfect valley filter and energy splitter controlled by electric gating. 
  By changing the relative electric field magnitude,  the intersection of three topological zero-lines can transform smoothly into a single zero line, and the current partition can be controlled precisely.
  We explore the available methods for modulating this device systematically by changing the Fermi energy, 
  the energy gap size, and the size of central gapless region. The current partition is also influenced by magnetic fields  and the system size.
  Our results provide a microscopic depiction of the electronic transport properties around 
  a unit cell of minimally twisted bilayer graphene  and have far-reaching implications in the design 
  of electron-beam splitters and interferometer devices.
\end{abstract}

\maketitle

\textit{Introduction---.} The triangular networks of topologically confined states in graphene-based moir\'e patterns, which arise naturally in minimally twisted bilayer graphene ({\textit{t}}-BG)~\cite{networkSTM2018, networkscience2018,networkAline2018,prbnetwork2018,networls-multi,network2013,science12,nethelin,networkwk,magneticfieldwk,sciencejun,PNAS} and graphene/hexagonal boron nitride ({\textit{h}}-BN) superlattices~\cite{pressure,GrBN}, have attracted much attention~\cite{networkSTM2018, networkscience2018,networkAline2018,prbnetwork2018,networls-multi,network2013,science12,nethelin,networkwk}. The presence of a minimally twist changes the local stacking order, arranging the AB/BA stacking regions periodically in space. The perpendicular electric field in {\textit{t}}-BG or the sublattice potential difference in a graphene/{\textit{h}}-BN bilayer leads to an energy gap that results in opposite valley Chern numbers at the AB/BA stacking regions. At the interfaces between different topological regions, topologically confined states (\textit{i.e.}, zero-line modes; ZLMs) appear and form networks~\cite{zlm,zlm1,folded,partition,kink,2013,2015,2016,NP,N2015,Natphys}. Recently, a network of topologically protected helical states was imaged in a minimally {\textit{t}}-BG through 
scanning tunneling microscopy~\cite{networkSTM2018,nethelin}, and optical techniques~\cite{science12}. 
Because of the three-fold rotational symmetry of the lattices, the elementary component of this network is the intersection of three ZLMs. However, the microscopic electronic transport properties of this intersecting point remains poorly understood.

In this Letter, we study the transport properties of three intersecting ZLMs connected to six terminals. When current is injected from one terminal, we find notable partitioning towards the forward channel, this situation is qualitatively different to the intersection of two ZLMs, 
where no forward propagation of current is observed. In the case of three intersecting ZLMs, the incoming current is partitioned towards the forward and the two adjacent zero-lines. 
This current partitioning depends strongly on the size of the central region when the region 
is small but saturates when it is large. 
By tuning the Fermi level, we find that the current partition can be controlled over a wide range. 
A perpendicular magnetic field can tune the currents propagating to the adjacent lines 
whereas the forward-propagating current remains quite robust. 
By changing the size of the AA stacking zone, we also study the effect of a twist angle on the transport properties. 
We find that (i) our device can support stable current partitioning even without a perpendicular electric field and 
(ii) the electric field allows to tune the partition properties and turn the device into a valley-current splitter.

\begin{figure}
	\includegraphics[width=8.6cm,angle=0]{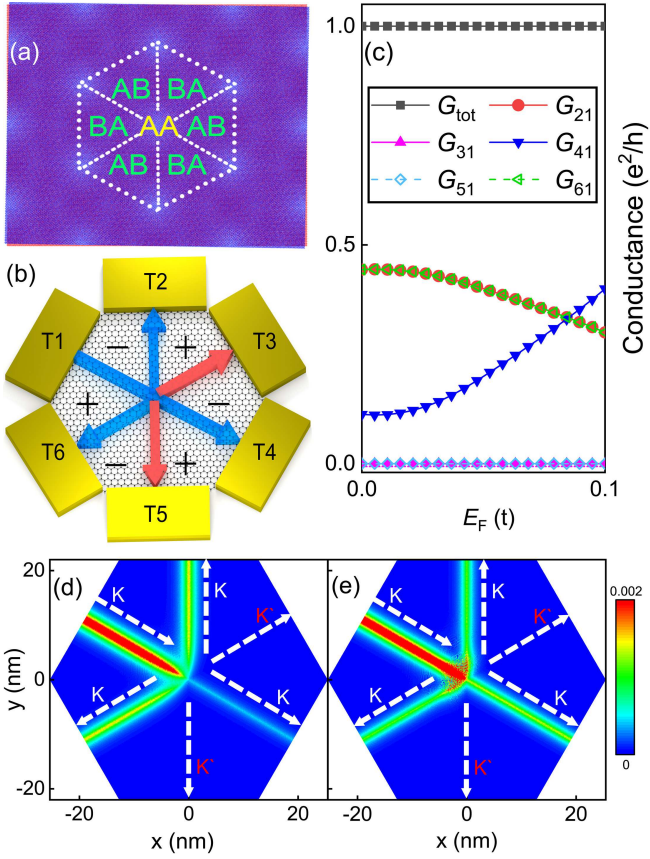}
	\caption{(a) Moir\'e pattern of twisted bilayer graphene and identification of regions with AB, BA, and AA stacking.
		(b) Schematic of six-terminal device with three intersecting zero-lines. 
		The $+$ and $-$ signs indicate the alternating sublattice potentials. Throughout this paper and unless specified otherwise, we use a side length of 25~nm for our regular hexagonal device. 
		The blue and red arrows correspond to modes that carry valley indices K and K$'$, respectively.
		Transmission is forbidden along the zero-lines with reversed chirality indicated by the red arrows.
		(c) Dependence of current partitioning as a function of Fermi level to the different output terminals. $G_{j1}$ is the conductance 
		between T1 to T$j$, while $G_{tot}$ is the total conductance between T1 to all the other terminals, 
		which remains quantized when the Fermi level shifts from the charge neutrality point and gradually approaches the bulk gap edge.
		(d) and (e) Local density of states of incoming current at Fermi level energies $10^{-3}$t and 0.1t, respectively. 
		The green arrows show the forward propagation directions of valley K currents, 
		while the white arrows represent the the zero-lines with valley index K$'$ where the currents do not propagate.
		The color bar shows a linear gradient of values.}
	\label{fig1}
\end{figure}

\textit{Model Hamiltonian---.}
The general stacking order of the moir\'e pattern is shown in Fig.~1(a), where the bright zone in the center corresponds to AA stacking. The size of AA stacking region decrease with the decreasing of the twist angle. 
Around this central zone, alternating periodic chiral AB or BA stacking regions are formed. At the interfaces between the AB and BA stacking regions, domain walls appear, indicated by the white dashed lines. By applying a perpendicular electric field, the centers of the chiral stacking regions become gapped while gapless ZLMs form at the domain walls~\cite{networkSTM2018,nethelin,science12}. The AA stacking zone is located at the intersection of three concurrent ZLMs, which are successively rotated by $60^{\circ}$. Without loss of generality, in our calculations we consider a monolayer graphene flake with position-dependent staggered sublattice potentials to form six adjacent regions with different valley Hall topologies labeled by $+/-$ signs in Fig.~\ref{fig1}(b), that correspond to AB/BA stacking regions under an electric field in a twisted bilayer graphene. The valley Hall domain walls form three intersecting zero-lines while the intersection region corresponding
to the AA stacking zone in the moir\'e pattern of a twisted bilayer is not gapped locally. 
The monolayer graphene flake with staggered site potentials can be described 
by the following $\pi$-orbital tight-binding Hamiltonian 
\begin{eqnarray}
 	H=-t \sum_{ \langle ij \rangle} c_i^{\dag} c_j+\sum_{ i\in A}U_{Ai}c_i^{\dag}c_j+\sum_{ i\in B}U_{Bi}c_i^{\dag}c_j, \nonumber
\end{eqnarray}
where $c_i^{\dag}(c_i)$ is a creation (annihilation) operator for an electron at site $i$, and $t=2.6$~eV is the nearest-neighbor hopping amplitude. The sublattice potentials are spatially varying, with $U_{Ai}$=$-U_{Bi}$=$\lambda\varDelta$ in regions labeled by $\lambda = \pm$ as shown in Fig.~1(b), where 2$\varDelta$ measures the magnitude of the staggered sublattice potential difference.

The zero-lines are connected to six reservoirs labeled T$_i$ ($i=1$--$6$) as shown in Fig.~\ref{fig1}(b). Herein, we take T$_1$ to be the injection terminal. The electronic transport calculations are based on the Landauer--B\"{u}ttiker formula~\cite{Datta} and recursively constructed Green's functions~\cite{wang}. The conductance from $q$-terminal to $p$-terminal is evaluated from
\begin{eqnarray}
  G_{pq}=\frac{2e^2}{h} {\rm Tr}[\Gamma_{p} G^r \Gamma_{q} G^a], \nonumber
\end{eqnarray}
where $G^{r,a}$ is the retarded/advanced Green's function of the central scattering region, and $\Gamma_{p}$ is the line-width function describing the coupling between $p$-terminal and the central scattering region. The propagation of a ZLM coming in from $p$-terminal is illustrated by the local density of states at energy $\epsilon$, which can be calculated by
\begin{eqnarray}
	\rho_p(r,\epsilon)=1/2\pi[G^r\Gamma_{p}G^a]_{rr}, \nonumber
\end{eqnarray}
where $r$ is the actual spatial coordinate.

\textit{Current Partition Laws---.} In our calculations, the central region is a hexagon with circumcircle diameter $D=50$~nm, and the diameter of AA stacking zone is $0.28~$nm unless stated otherwise. With $\varDelta=0.1t$, we calculate how the current partition depends on the Fermi energy $E_{\rm F}$ as shown in Fig.~1(c). The current partition laws can be summarized as:
\begin{eqnarray}
 	&&G_{31}=G_{51}=0, \label{eq1}\\
 	&&G_{21}=G_{61}, \label{eq2}\\
 	&&G_{tot}=G_{21}+G_{41}+G_{61}={e^2}/{h}. \label{eq3}
\end{eqnarray}

Here, Eq.~(\ref{eq1}) restricts current partitioning into zero-lines with opposite chirality, while the condition in 
Eq.~(\ref{eq2}) requires the mirror reflection symmetry of the partition, that is broken in the presence of a magnetic field as we show later. Equation~(\ref{eq3}) implies that there is no backscattering due to inter-valley scattering, and indeed any reflected current remains very weak even in the presence of disorders~\cite{zlm}.

Note that the forward-propagating conductance $G_{41}$ is nonzero because the zero-line from T$_1$ to T$_4$ has the same chirality, 
in contrast to the case of two intersecting zero-lines where the forward propagation is forbidden by the chirality conservation rule. 
Moreover, the forward-current transmission strongly depends on $E_{\rm F}$ as shown in Fig.~\ref{fig1}(c). 
When $E_{\rm F}$ is close to the charge-neutrality point (CNP), $G_{41}\approx 0.11 {e^2}/{h}$ is less than the conductance of $G_{21}=G_{61} \approx 0.44 {e^2}/{h}$ towards the sides. 
When $E_{\rm F}$ is shifted away from the CNP and moves toward the bulk band edges, $G_{41}$ increases gradually and exceeds $G_{21}$ when $E_F > 0.08 t$. The current partition at different $E_{\rm F}$ is shown more clearly in Figs.~\ref{fig1}(d) and (e), wherein the local density of states for current injected from T$_1$ is plotted at $E_{\rm F}=0.001t$ and $0.100t$, respectively. 
This Fermi-energy dependent current partition suggests that longitudinal transport  
may be greatly modified through a perpendicular electric gate by altering the current 
percolation properties through multiple partition nodes. 

\begin{figure}
  \includegraphics[width=8.6cm,angle=0]{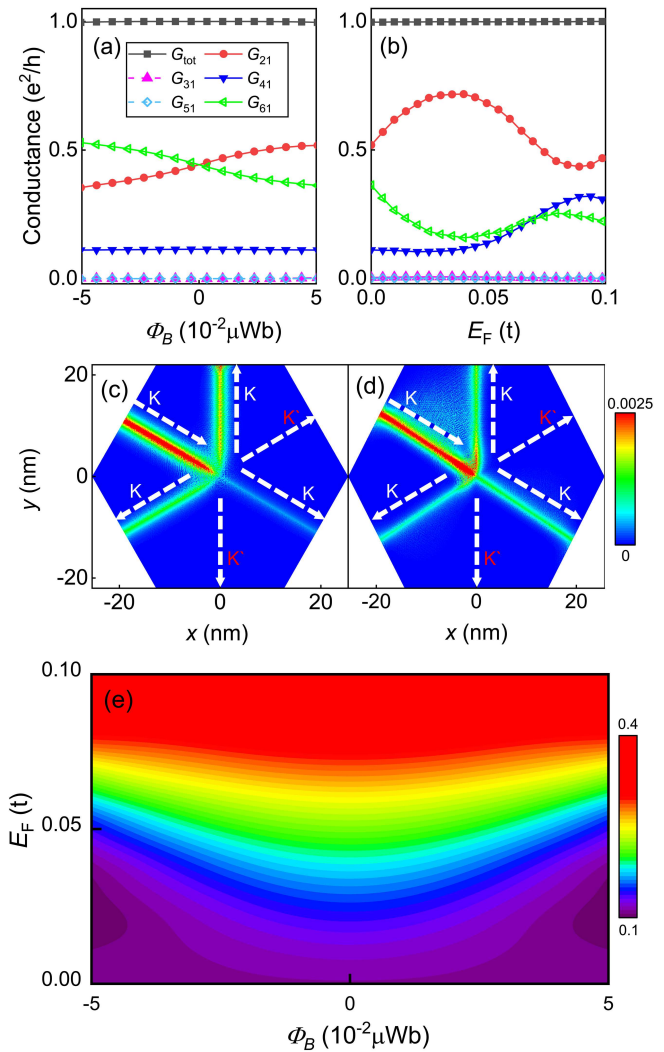}
  \caption{ (a) Conductance near charge-neutrality point $E_{\rm F}=0.001t$ for different values of magnetic flux $\varPhi_B$. (b) Conductance under the same magnetic flux $\varPhi_B$ for different values of Fermi energy $E_{\rm F}$.	(c) and (d) Local density of states of injected current for Fermi levels $10^{-3}t$ near the charge neutrality point and $0.1t$ near the bulk band edge with magnetic flux \SI{0.05}{\micro\weber} that illustrates the asymmetric partition of currents towards the side leads.
  	% {\bf Is it possible to label the figures the different values of the Fermi energy or indicate CNP and band edge? It might be useful to represent in units of $\Delta$.}
 (e) Phase diagram for conductance $G_{14}$ with different values of $E_{\rm F}$ and $\varPhi_B$ for $\varDelta=0.1t$.  The color bar shows a linear gradient of values in units of ${e^2}/{h}$. }
  \label{fig2}
\end{figure}

\textit{Influence of Magnetic Field---.} 
In addition to control through electrical means, current partition is also strongly affected by applying a magnetic field. 
The effect of a magnetic field $\mathbf{B}=\nabla\times \mathbf{A}$ can be included by attaching a 
Peierls phase factor to the following hopping term:
\begin{equation}
   t_{i,j}\longrightarrow t_{i,j}\exp(-i\frac{e}{\hbar}\int \mathbf{A} \cdot \mathrm{d}l), \nonumber
\end{equation}
where $\int \mathbf{A}\cdot \mathrm{d}l$ is the integral of the vector potential along the path from site $i$ to $j$. The presence of a magnetic field changes the current partition as shown in Fig.~\ref{fig2}(a) for $\varDelta=0.1t$ and $E_{\rm F}=0.001t$. We find that the presence of the magnetic field removes the equality between $G_{21}$ and $G_{61}$ because the magnetic field breaks the mirror reflection symmetry. 
However, the forward-propagating current remains unaffected.

By changing $E_{\rm F}$ under a given magnetic field as shown in Fig.~2(b), we find that the current partition to both forward and side directions vary simultaneously. As $E_{\rm F}$ approaches the bulk band edge, the current partitioned to T$_4$ increases gradually but does not reach the same magnitude as that for vanishing magnetic flux. Figures~\ref{fig2}(c) and (d) show more clearly the current partition under a magnetic flux of $\varPhi_B= \SI{0.05}{\micro\weber}$ for $E_{\rm F}=0.001t$ and $0.1t$, respectively, wherein more current is partitioned into T$_2$ than into T$_6$. This difference is more obvious at the higher Fermi energy, in agreement with previous work~\cite{networkwk, magneticfieldwk}.

To show the influences of magnetic field and Fermi energy more systematically, in Fig.~\ref{fig2}(e) we show the phase diagram for conductance $G_{41}$ with different values of $E_{\rm F}$ and $\Phi_B$. $G_{41}$ is only weakly dependent on the magnetic field as $E_{\rm F}$ approaches CNP and the bulk band edges but is strongly dependent on the magnetic field in the middle region.
\begin{figure}
  \includegraphics[width=8.6cm,angle=0]{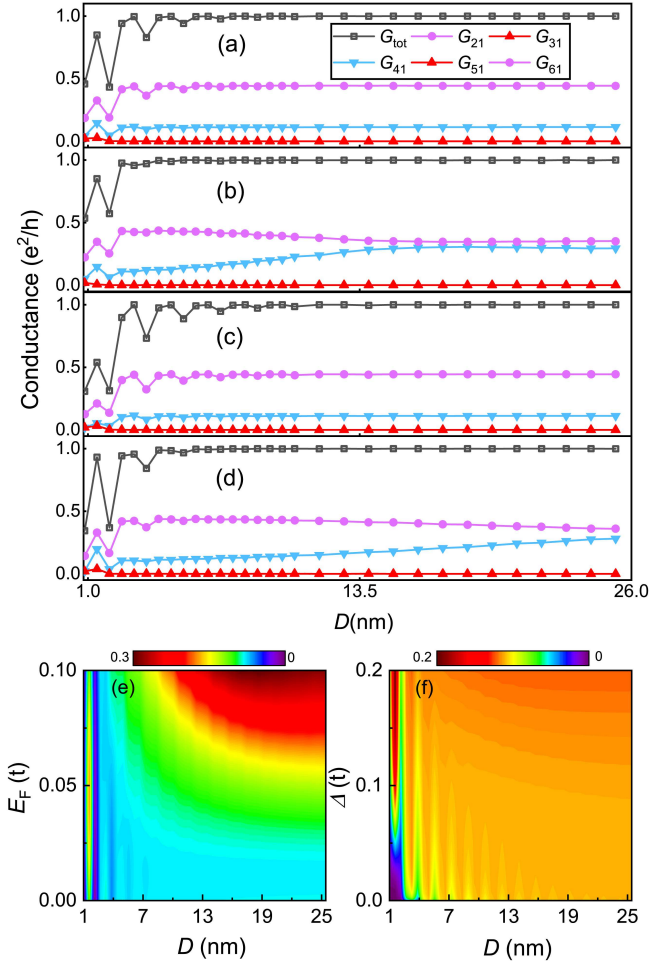}
  \caption{
%{\bf Can some of the labeling be done within the figure to make it easier for the reader?}
Dependence of current partitioning on device circumcircle diameter $D$ for different output leads with (a) potential $\varDelta=0.1t$ and Fermi energy $E_{\rm F}=0.001t$, (b)	$\varDelta=0.1t$ and $E_{\rm F}=0.1t$, (c) 	$\varDelta=0.05t$ and $E_{\rm F}=0.001t$, and (d) 	$\varDelta=0.05t$ and $E_{\rm F}=0.05t$.	Phase diagrams for conductance $G_{41}$ with (e) different values of $E_{\rm F}$ and $L$ but the same potential $\varDelta=0.1t$ and (f) different values of $\varDelta$ and $L$ but the same Fermi energy $E_{\rm F}=0.001t$. The color bar shows a linear gradient of values in units of $e^2/h$. } %$\frac{e^2}{h}$.}
  \label{fig3}
\end{figure}

\textit{Effect of System Size---.} For an electronic nanodevice, its size plays a crucial role. Herein, we investigate the current partition of a device whose circumcircle diameter $D$ ranges from 1~nm to 25~nm. In Figs.~\ref{fig3}(a)--(d), we plot the current partition as a function of $D$ at different Fermi energies $E_{\rm F}$ and staggered sublattice potential $\varDelta$. The current partition fluctuates obviously for $D<6$~nm, which is mainly due to the backscattering because the total conductance is not quantized. When the system size $L$ increases, the backscattering vanishes. However, increasing the system size has only a weak influence on the current partition when $E_{\rm F}$ is close to CNP but has a definite influence to bring $G_{41}$ and $G_{21}$ closer together when $E_{\rm F}$ lies at the band edge. Such a behavior suggests that the equal-current partition approximation is valid at small twist angle but invalid at large twist 
angle when the Fermi energy $E_{\rm F}$ is far away from charge neutrality. 

% It would be great to come up with an intuitive explanation of the calculations beyond the descriptive text.
The length dependence of the current partition at different $E_{\rm F}$ with $\varDelta=0.1t$ is shown in Fig.~\ref{fig3}(e), 
wherein the fluctuation appears for small sizes and exhibits weak dependence on $E_{\rm F}$. 
As the system size increases, $G_{41}$ gradually increases and then saturates at a magnitude that increases with $E_{\rm F}$. 
By setting $E_{\rm F}=0.001t$, we plot the phase diagram of $G_{41}$ as function of the energy gap $\varDelta$ and circumcircle diameter $D$ in Fig.~\ref{fig3}(f), wherein similar behaviors are observed. Moreover, we find that the saturated $G_{41}$ at larger size also increases with the energy gap.
\begin{figure}
	\includegraphics[width=8.6cm,angle=0]{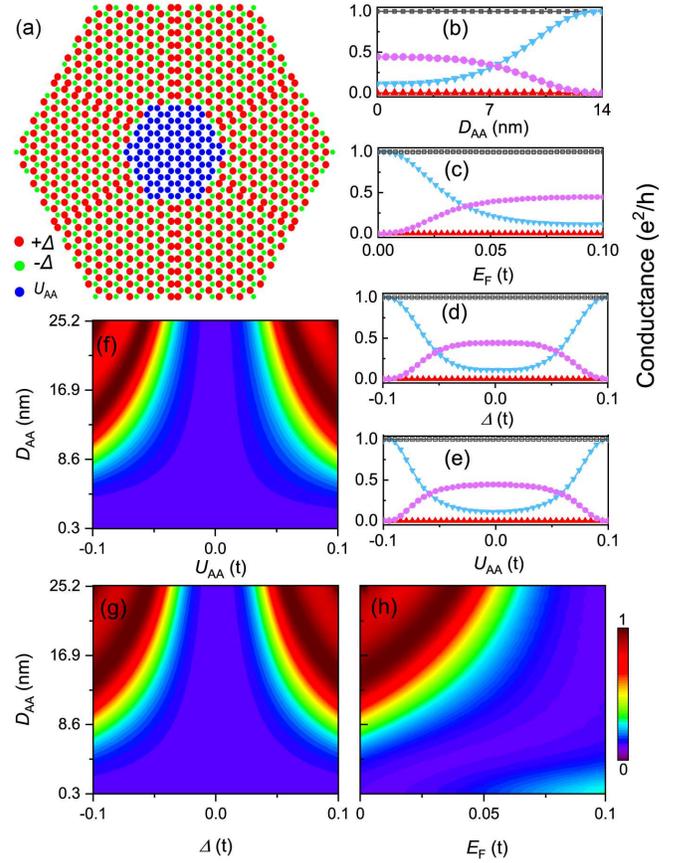}
	\caption{(a) Schematic of lattice structure. The red (green) points are lattice points with positive (negative) potential $\varDelta$, and the blue points are lattice points with potential $U_{\rm AA}$.
		(b)$-$(e) Dependence of the current partition into different outgoing leads (describe the curves) for different sampling 
		of system parameters such as the diameter $D_{\rm AA}$ of AA stacked zone, 
		the Fermi energy $E_{\rm F}$, and staggering potentials $\varDelta=U_{\rm AA}$, represented as a function of 
		(b) $D_{\rm AA}$ for $E_{\rm F}=0.001t$ and $\varDelta=U_{\rm AA}=0.1t$, 
		(c) $E_{\rm F}$ for $D_{\rm AA}=14$~nm and $\varDelta=U_{\rm AA}=0.1t$,
		(d) $\varDelta$ for $D_{\rm AA}$~nm, $E_{\rm F}=0.001t$, and $U_{\rm AA} = \varDelta$,   %{\bf details are missing}
		(e) $U_{\rm AA}$ for $D_{\rm AA}=14$~nm, $E_{\rm F}=0.001t$, and $\varDelta=0.1t$.
		%		(b) Dependence of current partitioning on diameter $D_{\rm AA}$ of AA stacked zone for different outgoing leads for Fermi energy $E_{\rm F}=0.001t$ and potentials $\varDelta=U_{\rm AA}=0.1t$.
		%		(c) Dependence of current partitioning on $E_{\rm F}$ for different outgoing leads for $D_{\rm AA}=14$~nm,  and potentials $\varDelta=U_{\rm AA}=0.1t$.
		%		(d) Dependence of current partitioning on $\varDelta$ for different outgoing leads for $D_{\rm AA}$~nm, $E_{\rm F}=0.001t$, and $U_{\rm AA} = \varDelta$.
		%		(e) Dependence of current partitioning on $U_{\rm AA}$ for different outgoing leads for $D_{\rm AA}=14$~nm, $E_{\rm F}=0.001t$, and $\varDelta=0.1t$.
		(f) Phase diagram for conductance $G_{41}$ with different values of $U_{\rm AA}$ and $D_{\rm AA}$ for $E_{\rm F}=0.001t$ and $\varDelta=0.1t$.
		(g) Phase diagram for conductance $G_{41}$ with different values of $\varDelta$ and $D_{\rm AA}$ for $E_{\rm F}=0.001t$ and $U_{\rm AA} = \varDelta$.
		(h) Phase diagram for conductance $G_{41}$ with different values of $E_{\rm F}$ and $D_{\rm AA}$ for $\varDelta=U_{\rm AA}=0.1t$.
		The color bar shows a linear gradient of values in units of $e^2/h$, %$\frac{e^2}{h}$, 
		and the current-partitioning labels are the same as in Fig.~3. }
	\label{fig4}
\end{figure}

\textit{Effect of AA Stacking Size---.} We now study how the AA stacking size at the intersection affects the current partition, as shown schematically in Fig.~4(a). This can simulate the effect of the twisting angle $\theta$ in $\textit{t}$-BG, where the size of AA stacking zone shrinks for smaller twist angle $\theta$~\cite{Morie}. We vary the circumcircle diameter of the AA stacking region, namely $D_{\rm AA}$, from the smallest size to 25.2~nm while fixing the device size as $D=50~$nm. With $E_{\rm F}=0.001t$ and $\varDelta=0.1t$, we find that $G_{41}$ increases with $D_{\rm AA}$ and becomes quantized at $D_{\rm AA} \approx 14~$nm, where the current partition of $G_{21, 61}$ vanishes. This behavior suggests that the central AA stacking region acts as a scattering zone whose scattering weakens as its size increases, thereby increasing the forward-propagation current. As a result, most of the current flows along one zero-line, with weak partition to the other allowed zero-lines.

For the largest AA stacking region, i.e., when there is no longer an energy gap in the central scattering region, the current is fully partitioned into the forward-propagating zero-line. By changing $E_{\rm F}$ from CNP, we find that the current partition becomes tunable. As we increase $E_{\rm F}$, $G_{41}$ first decreases and then increases to the quantized value near $E_{\rm F} \approx 0.05t$. When we further increase $E_{\rm F}$, $G_{41}$ decreases to a small value. For $D_{\rm AA}=14$~nm and $\varDelta=0.1t$, the current partitioned into $L_4$ grows with $|\varDelta|$ and those into $L_2$ and $L_6$ decrease.

Note that all of the above results were calculated under the condition that $U_{\rm AA} = \varDelta$. For $E_{\rm F}=0.001t$ and $\varDelta=0.1t$, we find that $G_{41}$ increases with $|U_{\rm AA}|$ and becomes quantized at $|U_{\rm AA}|=\varDelta$, where the current partitioning of $G_{21, 61}$ vanishes.

In Fig.~4(f), we plot the phase diagram of $G_{41}$ with different $U_{AA}$ and $D_{\rm AA}$ at fixed $E_{\rm F}=0.001t$ and $\varDelta=0.1t$; for $D_{\rm AA}<4.3~nm$, $U_{AA}$ no longer influences the current partitioning. In Fig.~4(g), we also plot the phase diagram of $G_{41}$ with different $\varDelta$ and $D_{\rm AA}$ at fixed $E_{\rm F}=0.001t$. Now, the resonant transmission appears for a different band gap $\varDelta$. As we decrease $|\varDelta|$ from $0.1t$, the value of $D_{\rm AA}$ for resonant transmission increases until it equals 25.2 nm, and the resonant transmission disappears when $|\varDelta|< 0.6t$. Figure~\ref{fig4}(h) plots the phase diagram of $G_{41}$ with different $E_{\rm F}$ and $D_{\rm AA}$ at fixed $\varDelta=0.1t$. One finds that the resonant transmission appears at $E_{\rm F} < 0.05t$ and the resonant diameter increases with $E_{\rm F}$. Note that at the resonant transmission point of $\varDelta=0.1t$ and $D_{\rm AA} \approx 14~$nm, increasing $E_{\rm F}$ drives $G_{41}$ to decrease gradually as in Fig.~4(c), suggesting that the current partition goes from zero to a finite value. 
Such tunability by means of electric gating could be used to construct a field-effect transistor based on the dissipationless 
topological ZLMs for switching on and off the forward propagation. 

\textit{Tunable valley-current splitter---}
In order to control the current partition effectively and quantitatively, we add two set of electric field $\Delta_1$ and $\Delta_2$ on the six terminal device, as shown in the insert of Fig.~5(a).
We vary the magnitude of $\Delta_2/\Delta_1$ while fixing the $E_{\rm F}=0.001t$ , device size as $D=50 nm$ and $\Delta_1=0.1t$. We find that $G_{41}$ becomes quantized when $\Delta_2/\Delta_1<0$, where the current partition of $G_{21,61}$ vanishes, which means there only exit a single zero-line.  With $\Delta_2/\Delta_1>0$, we find that $G_{41}$ decreases and  $G_{21,61}$ increase with $\Delta_2/\Delta_1$.  When $\Delta_2/\Delta_1\approx0.6$, $G_{41,21,61}$ become the same value, where present the intersection of three zero-lines. When we further increase $\Delta_2/\Delta_1$ to 2, $G_{41}$ decrease to zero and $G_{21,61}$ become 0.5 $\rm e^2/h$, as shown in Fig.~5(c), where shows the current partition just like the intersection of two zero-lines. By fixing one of the electric field and vary the other, we can modulate the partition $G_{41}$ from zero to quantized, precisely, making this device more qualified as a valley-current splitter. 

\begin{figure}
	\includegraphics[width=8.6cm,angle=0]{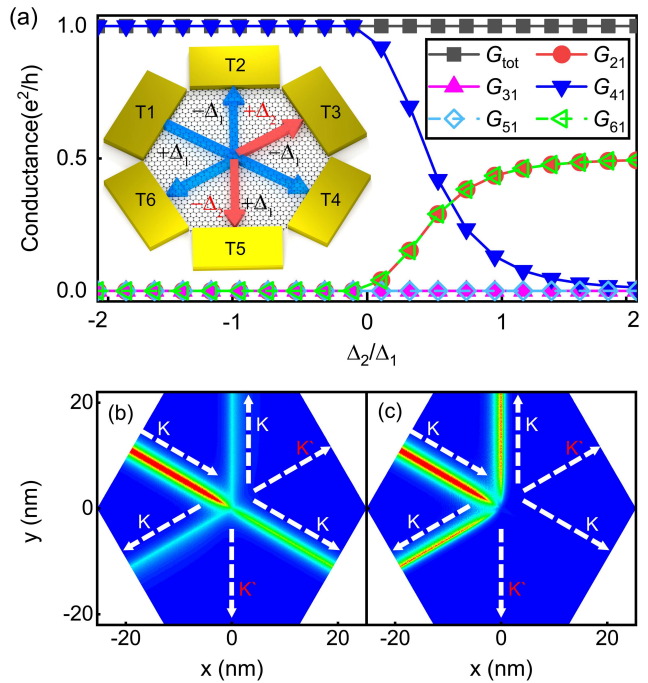}
	\caption{ 
		(a) Dependence of current partitioning as a function of $\Delta_2/\Delta_1$ for different output leads with Fermi energy $E_{\rm F}=0.001t$. Schematic of six-terminal device with two-pair of gates labeled by $\Delta_1$ and $\Delta_2$ is in the insert.
	   (b) and (c) Local density of states of incoming current at $\Delta_2/\Delta_1=0.2$ and $\Delta_2/\Delta_1=2$, respectively, at the same Fermi level energies $0.001t$.  
    }
	\label{fig5}
\end{figure}

\textit{Summary---.} We presented a systematic study of the electronic transport properties of three intersecting zero-lines in a twisted bilayer graphene, which form six topological zones stacked either in AB or BA patters and is centered in a gapless region with AA stacking. We investigated the influence of system size, AA-region size, $E_{\rm F}$, energy gap, and magnetic field on the current partitioning. When current flows through the device, the forward-propagating current partitioning $G_{41}$ is nonzero and can be tuned by mediating $E_{\rm F}$. This is in contrast to the topological intersection of two zero-lines, where forward propagation is forbidden.

When the central AA region is small, increasing $E_{\rm F}$ to the band edge enhances $G_{41}$ to partition the current into three nearly equal zero-lines, and the potential of the AA region does not affect the current partitioning. By making the central AA region larger, which corresponds to decreasing the twist angle, resonant forward transmission can be realized at a proper $E_{\rm F}$, in which case the partitioning to the side zero-lines vanishes. Starting from the case of resonant transmission, further increasing $E_{\rm F}$ (e.g., by means of electric gating) increases the current partitioning to the side zero-lines from zero to a definite level, thereby suggesting that the system could be used as a dissipationless field-effect transistor. Moreover, decreasing the system size gives rise to strong backscattering and conductance fluctuations. In the absence of a magnetic field, the symmetric geometry of our central region leads to $G_{21}=G_{61}$ current partitioning, but the presence of a magnetic field breaks this symmetry.

Our theoretically proposed device can pericely find its experimental realization in graphene moir\'e structures, and can also find realizations in phononic crystals. Specifically, our findings are the first prediction of the current partition at the zero-line 
intersection node of a triangular network of topological channels in small twist angle bilayer 
graphene and paves the way towards the realization of low-power topological quantum devices.

\begin{acknowledgments}
\textit{Acknowledgments---.} This work was supported financially by the National Key R \& D Program (2017YFB0405703 and 2016YFA0301700), and the NNSFC (11474265, and 11674024), the China Government Youth 1000-Plan Talent Program, and Anhui Initiative in Quantum Information Technologies. We are grateful to AMHPC and Supercomputing Center of USTC for providing high-performance computing assistance.
\end{acknowledgments}

\end{document}